\documentclass[11pt]{article}
\usepackage{amsmath,amssymb,latexsym,graphicx,psfrag}
\newcommand{\sect}[1]{\setcounter{equation}{0}\section{#1}}
\newcommand{\subsect}[1]{\subsection{#1}}

\newfont{\extra}{msbm10 scaled\magstep1}

\newcommand{\extr}[1]{\mbox{\extra #1}}

\def\d(#1){\partial_{#1}}

\def\R{\extr R}

\def\Z{\extr Z}

\def\be{\begin{equation}}
\def\ee{\end{equation}}
\def\bea{\begin{eqnarray}}
\def\eea{\end{eqnarray}}

\def\p{\partial}
\def\a{\alpha}

\def\l{\lambda}

\begin{document}

\begin{center}{ \LARGE \bf
Dynamical symmetries \\[2ex]
for superintegrable  quantum systems}
\end{center}
\vskip0.25cm

\begin{center}
J.A. Calzada~$^1$, J. Negro~$^2$, M.A. del Olmo~$^2$ 
\vskip0.25cm

{\it ${}^1$~Departamento de Matem\'atica Aplicada,\\
${}^2$~Departamento de  F\'{\i}sica Te\'orica,\\
 Universidad de  Valladolid,
 E-47011, Valladolid,  Spain}
\vskip0.15cm

E. mail:
{juacal@eis.uva.es}, {jnegro@fta.uva.es}, {olmo@fta.uva.es}
\end{center}

\vskip1.5cm 
\vskip0.5cm

\begin{abstract}
We study the dynamical symmetries of a class of two-dimensional superintegrable systems on a 2-sphere, obtained by a  procedure based on the Marsden-Weinstein reduction,  by  considering  its shape-invariant intertwining operators. These are obtained by  generalizing the techniques of factorization of one-dimensional systems. 
We firstly obtain a pair of noncommuting Lie algebras $su(2)$ that originate the  algebra $so(4)$. By considering  three spherical coordinate  systems we get the algebra $u(3)$  that can be enlarged by `reflexions' to $so(6)$.  The bounded eigenstates of the Hamiltonian hierarchies are associated to the irreducible unitary representations of these dynamical algebras.  
\end{abstract}
\sect{Introduction}

It is well known that integrable Hamiltonian systems play a fundamental role in the
 description of physical systems, because  their
  many interesting properties both from mathematical and physical points of
view. Many of these Hamiltonian systems have proved to be of an extraordinary
physical interest e.g. harmonic oscilator, Kepler problem, Morse \cite{Morse}, Posch--Teller \cite{Posch}, Smorodinski-Winternitz~\cite{smorodinski, Evans1}, Calogero \cite{Calogero1} and
Sutherland \cite{Sutherland} potentials.

In this paper we present a class of integrable Hamiltonian systems that allow us to 
generalize  the intertwining  transformations for one-dimensional (1D) systems \cite{infeld} to 
higher dimensions.

These Hamiltonian systems are superintegrable, i.e.,  they have more than N 
constants of motion, being N the dimension of the configuration space for the Hamiltonian system. Although these motion integrals are  not  all of  them in involution they determine more than one subset of N constants (in all the cases one of them is the Hamiltonian) in involution. 
The system is said to be superintegrable in the sense of \cite{Eisenhart1} or maximally superintegrable if there exist  $2\,N - 1$ invariants well defined in phase--space.

Using  the Marsden--Weinstein reduction procedure \cite{Marsden} we
construct such classical systems starting from a free Hamiltonian
lying in an N-dimensional homogeneous space of a
suitable Lie group, whose  action allows to us to calculate a momentum map that assure the integrability, or even superintegrability of the reduced systems
\cite{Evans1}. From an opposite point of view, there are good reasons to suspect
that any integrable system may be constructed as a reduction from a
free one \cite{Grabowski}.

For the corresponding quantum systems we present here a generalization to higher dimensional spaces of the intertwining  transformations for 1D factorizable systems. These 1D systems have  dynamical Lie
algebras of rank one generated by the intertwining operators \cite{infeld}.  By using a concrete superintegrable Hamiltonian system with underlying symmetry the Lie algebra $su(3)$ we find that its dynamical symmetry can bee enlarged to $so(6)$. However, these results can be implemented to higher dimensional systems of the same class, and also can be helpful   in the study of  other kinds of integrable systems  using  algebraic methods \cite{kuru1,kuru2,samani,ranada,santander,ioffe}.

 In sections~\ref{SU(p,q)hamiltoniansystems}, 
 \ref{aclassicalsuperintegrableu(3)hamiltoniansystems} and 
 \ref{theHamilton-Jacobiequation}
we will introduce a two-dimensional superintegrable system and find some separable solutions by standard procedures (Hamilton-Jacobi equation).  Next, in  
sections~\ref{aquantumsuperintegrableu(3)hamiltoniansystem} and 
\ref{dynamicalsymmetries} we study the corresponding Schr\"odinger equation which is factorizable in two 1D equations. We construct some sets of intertwining operators closing  the Lie algebras $u(3)$ and $so(6)$ by taking also into account different separable coordinate systems as well as  discrete symmetries. We characterize the eigenfunctions of the Hamiltonian hierachies, obtained from the intertwining operators, as irreducible unitary 
representations of these dynamical algebras.  

\sect{Superintegrable $SU(p,q)$-Hamiltonian systems} 
\label{SU(p,q)hamiltoniansystems}

Let us consider a free Hamiltonian 
$H=\,4\,g^{\mu\bar{\nu}}\,p_\mu\,\bar{p}_\nu$ 
($\mu,\nu=0\,,\cdots,n=\,p+q-1$ and the bar stands for complex conjugate) defined in the configuration space 
 $    \frac{SU(p,q)}{SU(p-1,q)\times U(1)}$. This is an hermitian hyperbolic space with metric  $g_{{\mu}\nu}$ and coordinates 
$y^\mu\in \mathbb{C}$, verifying $g_{\bar{\mu}\nu}\,\bar{y}^\mu\,y^\nu=\,1$ (by 
$p_\mu$ we denote the conjugate momenta).  The geometry and properties of this kind of spaces  are described in \cite{Kobayashi,Calzada1}.  

Using a maximal abelian subalgebra (MASA) of the Lie algebra $su(p,q)$ \cite{Olmo1} the reduction procedure allows us to obtain a reduced Hamiltonian,
 $ H=\,\frac{1}{2} g^{\mu \nu}\,p_{s^\mu}\,p_{s^\nu} + V(s)$,  lying in the corresponding reduced space  (a homogeneous $SO(p,q)$-space),
  where $V(s)$ is a potential depending on the real coordinates
$s^\mu$ satisfying the constraint  $g_{{\mu}\nu}\,{s}^\mu\,s^\nu=\,1$. 

The set of complex coordinates $y^\mu$ is transformed by
the reduction into a set of ignorable variables $x^\mu$ (which 
are the parameters of the transformation associated to the 
MASA of $u(p,q)$ used in the
reduction) and the actual real coordinates $s^\mu$.

If $Y_\mu\,,\mu=\,0,\cdots,n$, is a basis of the considered MASA of
$u(p,q)$, formed by pure imaginary matrices (this is a basic hypotesis in our reduction procedure), the relation between
old ($y^\mu$) and  new coordinates ($x^\mu\,,s^\mu$) is 
\begin{equation*}
    y^\mu=\,B(x)^{\mu}_{\, \nu}\,s^\nu\,,\qquad
    B(x)=\,\text{exp}\left(x^\mu\,Y_\mu\right)\,.
    \end{equation*}
This relationship assures the ignorability of the $x$ coordinates (the vector
fields corresponding to the MASA are straightened out in these
coordinates). The Jacobian matrix, $J$, of the coordinate transformation is given explicitly  by
\begin{equation*}
    J=\,\frac{\partial(y\,,\bar{y})}{\partial (x\,,s)}=\,
    \begin{pmatrix}
        A & B \\
        \bar{A} & \bar{B}
    \end{pmatrix}\,, \qquad
    A_{\,\nu}^{\mu} =\,\frac{\partial y^\mu}{\partial
    x^\nu}=\,\left(Y_\nu\right)^{\mu}_{\,\rho}\,y^\rho\,.
\end{equation*}

The Hamiltonian calculated in the new coordinates is written as
\begin{equation*}
    H=\,c \,\left(\frac{1}{2}\,g^{\mu\,\nu}p_\mu \,p_\nu +
    V(s)\right)\,,\qquad V(s)=\,p_{x}^{T}\,(A^{\dagger}\,K\,A)^{-1}\,p_x\,,
\end{equation*}
where $p_x$ are the constant momenta  associated to the ignorable
coordinates $x$ and $K$ is the matrix defined by the metric $g$. A detailed exposition of this construction procedure of this family of superintegrable systems can be found in \cite{winternitz}.

\sect{A classical superintegrable $u(3)$-Hamiltonian system}
 \label{aclassicalsuperintegrableu(3)hamiltoniansystems}

To obtain the classical superintegrable Hamiltonian associated to the unitary
Lie algebra $su(3)$ using the
reduction procedure sketched in the previous section we are going to proceed in the following way \cite{olmo99}. 

Let us consider the  basis of $su(3)$  determined by   $3 \times 3$ matrices
$X_1\,,\cdots,X_8$, whose explicit form, when using the metric  $K=\text{diag}(1,1,1)$, is
\begin{alignat*}{4}
    X_1 =& \begin{pmatrix} i & 0 & 0\\ 0 &-i & 0\\ 0 & 0 &0\end{pmatrix}&\quad
    X_2 =& \begin{pmatrix} 0 & 0 & 0\\ 0 &i & 0\\ 0 & 0 &-i\end{pmatrix}&\quad
    X_3 =& \begin{pmatrix} 0 & 1 & 0\\ -1 &0 & 0\\ 0 & 0 &0\end{pmatrix}&\quad
    X_4 =& \begin{pmatrix} 0 & i & 0\\ i &0 & 0\\ 0 &
    0&0\end{pmatrix}\\[0.5cm]
    X_5 =& \begin{pmatrix} 0 & 0 & 1\\ 0 &0 & 0\\ -1 & 0 &0\end{pmatrix}&\quad
    X_6 =& \begin{pmatrix} 0 & 0 & i\\ 0 &0 & 0\\ i & 0 &0\end{pmatrix}&\quad
    X_7 =& \begin{pmatrix} 0 & 0 & 0\\ 0 &0 & 1\\ 0 & -1 &0\end{pmatrix}&\quad
    X_8 =& \begin{pmatrix} 0 & 0 & 0\\ 0 &0 & i\\ 0 & i&0\end{pmatrix}\,.
\end{alignat*}

In the compact case, as here with $su(3)$, there is only one MASA. This is  the Cartan subalgebra,
generated by the matrices
${\rm diag}(i,-i,0)$ and  ${\rm diag}(0,i,-i)$. 
However,  we shall work, in order to  facilitate the computations, with  the algebra $u(3)$ 
instead of $su(3)$. Hence, we  shall use the following basis for the corresponding MASA in $u(3)$
\[
    Y_0= {\rm diag}(i,0,0),\qquad
      Y_1= {\rm diag}(0,i,0),\qquad
    Y_2= {\rm diag}(0,0,i).
    \]
    
The actual real coordinates $s$ are related to the complex coordinates $y$ by 
 $y_\mu=\,s_\mu\,e^{i\,x_\mu}\; (\mu=0,1,2)$.
The Hamiltonian can be written as
\begin{equation} \label{hamiltonianou3}
    H =\,\frac{1}{2} \left(p_{0}^2 + p_{1}^2+p_{2}^2\right) + V(s)\,,\qquad 
    V(s)=\,\frac{m_{0}^2}{s_{0}^2} +
    \frac{m_{1}^2}{s_{1}^2} +\frac{m_{2}^2}{s_{2}^2} \,,
\end{equation}
lying in the 2-sphere  $ (s_0)^2+  (s_1)^2 +  (s_2)^2 = 1$.

To see that the system, so obtained, is superintegrable it is necessary 
to construct its invariants of motion. In this case we obtain three invariants
\be\label{invariantes}
  R_{\mu\nu}=\, (s_\mu\,p_\nu - s_\nu\,p_\mu)^2 + \left(m_\mu \frac{s_\nu}{s_\mu} +
        m_\nu \,\frac{s_\mu}{s_\nu}\,\right)^2 ,\qquad \mu<\nu, \; \mu=0,1,\;\nu=1,2 .
\ee
Note that only two of them are  in
involution at the same time (being one of them the Hamiltonian), so
the system is superintegrable in the sense of \cite{Eisenhart1}. 
The sum of these invariants is the Hamiltonian up to an additive
constant. 

\sect{The Hamilton-Jacobi equation for the $u(3)$-system}
 \label{theHamilton-Jacobiequation}

The  solutions of the motion problem for this system can be obtained  solving 
 the corresponding Hamilton--Jacobi (HJ)
equation in an appropriate coordinate system, such that the Hamilton--Jacobi
equation separates into a system of ordinary differential equations.

The 2-sphere 
can be  parametrized  on spherical coordinates
 $(\phi_1,\phi_2)$ around the $s_2$ axis \cite{Calzada1,Boyer}   by
\begin{equation}\label{coo}
   s_0 = \cos \phi_2\,\cos \phi_1,\qquad
   s_1 = \cos \phi_2\, \sin \phi_1,\qquad
   s_2 \,= \sin\phi_2
\end{equation}
   where $\phi_1\in [0,2 \pi)$ and  $\phi_2\in [\pi/2,3 \pi/2]$.
Then, the Hamiltonian (\ref{hamiltonianou3}) is written as
\begin{align*}
    H=&\,\frac{1}{2}\,\left(p_{\phi_2}^2 +
        \frac{p_{\phi_1}^2}{\cos^2\phi_2}\right) +
        V(\phi_1,\phi_2),\\[0.3cm]
    V(\phi_1,\phi_2)=&\,\frac{1}{\cos^2
        \phi_2}\,\left(\frac{m_{0}^2}{\cos^2\phi_1} +
        \frac{m_{1}^2}{\sin^2 \phi_1} \right) + \frac{m_{2}^2}{\sin^2
        \phi_2}\,.
\end{align*}
The potential is periodic and has singularities
along the coordinate lines 
$\phi_1=0,\pi/2,\pi,3\pi/2$,  and
$\phi_2=\pi/2,3\pi/2$, and
thre is  a unique minimum inside each
regularity domain. 

The invariants  (\ref{invariantes}) (denoted by
$\hat{I}$) can be written in terms of
the basis $\{X_1\,,\cdots\,,X_8\}$ as
$    \hat{I}_1 =\,X_{3}^2 + X_{4}^2\,,
    \hat{I}_2 =\,X_{5}^2 + X_{6}^2$ and
    $\hat{I}_3 =\,X_{7}^2 + X_{8}^2$.
The $u(3)$-Casimir is
\begin{equation*}
    C=\,4\,X_{1}^2 + 2\,\{X_1\,,X_2\} + 4\,X_{2}^2 + 3\,\hat{I}_1 + 3\,\hat{I}_2 +
    3\,\hat{I}_3\,,
\end{equation*}
where
the three first terms
are the second order operators in the enveloping algebra of the
compact Cartan subalgebra of $u(3)$.

The Hamiltonian is rewritten as
$H=\,I_1 + I_2 + I_3 + \text{constant}$, 
where by $I_i$ we denote the invariant $\hat{I}_i$  but rewritten in spherical
coordinates. So, we have
\begin{align*}
    I_1=&\,\frac{1}{2}\,p_{\phi_1}^2 + \frac{m_{0}^2}{\cos^2\phi_1} +
        \frac{m_{1}2}{\sin^2\phi_1},\\[0.3cm]
    I_2=&\,\tan^2\phi_2\,\left(\frac{1}{2}p_{\phi_1}^2
        \,\sin^2\phi_1 + \frac{m_{0}^2}{\cos^2\phi_1}\right) +
        \cos^2\phi_1\,\left(\frac{1}{2}\,p_{\phi_2}^2 +
        \frac{m_{2}^2}{\tan^2\phi_2}\right) \\[0.2cm]
       &\qquad + \, \frac{1}{2}\,p_{\phi_1}\,p_{\phi_2}\,\sin 2\,\phi_1\,\tan \phi_2 ,\\[0.3cm]
    I_3=&\, \tan^2\phi_2\,\left(\frac{1}{2}p_{\phi_1}^2
        \,\cos^2\phi_1 + \frac{m_{1}^2}{\sin^2\phi_1}\right) +
        \sin^2\phi_1\,\left(\frac{1}{2}\,p_{\phi_2}^2 +
        \frac{m_{2}^2}{\tan^2\phi_2}\right)\\[0.2cm]
    &\qquad - \frac{1}{2}\,p_{\phi_1}\,p_{\phi_2}\,\sin 2\,\phi_1\,\tan \phi_2 .
\end{align*}

Now, the HJ equation takes the form
\[
{1\over 2} \left( {\partial S\over \partial \phi_2}\right)^2
 +  {m_2^2\over\sin^2\phi_2}+
{1\over \cos^2\phi_2}\left({1\over 2}  \left( {\partial S\over
\partial \phi_1}\right)^2
+{m_0^2\over \cos^2\phi_1} + {m_1^2\over
\sin^2\phi_1}\right)=E .
\]
It separates into two ordinary
differential equations taking into account that the solution of the HJ equation can be written as
$S(\phi_1,\phi_2)=S_1(\phi_1)+ S_2(\phi_2)-Et$. Thus,
\begin{eqnarray*}
{1\over 2} \left( {\partial
S_1\over \partial
\phi_1}\right)^2+{m_0^2\over
\cos^2\phi_1} + {m_1^2\over
\sin^2\phi_1} & = & \alpha_1,\label{hjsu3}\\[0.3cm]
{1\over 2} \left( {\partial S_2\over \partial
\phi_2}\right)^2 +  {m_2^2\over
\sin^2\phi_2}+  {\alpha_1\over \cos^2\phi_2}
& = & \alpha_2 ,
\end{eqnarray*}
where $\alpha_2=E$ and $\alpha_1$ are the
separation constants (which are positive).
Each one of these two equations have the same form  
of those  corresponding to the 1D problem \cite{olmo99}.
 
 The solutions of both HJ equations are easily computed
and can be found as particular cases in
Ref.~\cite{Calzada1}.  A detailed  analysis of them
 shows that all the orbits in a
neighborhood of a critical point (center)
are closed and thus, the corresponding
trajectories are periodic (a direct consequence
of the correspondence between extrema of the
potential and critical points of the phase
space). 

 The explicit
solutions, when we restrict us to the domain 
$0<\phi_1,\phi_2<\pi/2$, are
\begin{eqnarray*}
\cos^2\phi_2 & = & {1\over 2E}\left[
b_2+\sqrt{b_2^2-4\alpha_1 E}\cos
2\sqrt{2E}t\right], \\
\cos^2\phi_1 & =  &{1\over 2\alpha_1}\left[
b_1+{1\over\cos^2\phi_2}
\left[{b_1^2-4\alpha_1 m_0^2\over 
b_2^2-4\alpha_1E}\right]^{1/2}\left(
(b_2\cos^2\phi_2-2\alpha_1)\sin
2\sqrt{2\alpha_1}\beta_1\right.\right.
\nonumber\\ 
&&\quad\left.\left.
+2\sqrt{\alpha_1}[(b_2
-E\cos^2\phi_2)\cos^2\phi_2-\alpha_1]^{1/2}
\cos2\sqrt{2\alpha_1}\beta_1\right)\right] ,
\end{eqnarray*}
where $b_1=\alpha_1+m_0^2-m_1^2$ and
$b_2=E+\alpha_1-m_2^2$.

Note that in this domain for the variables $(\phi_1,\phi_2)$ the minimum for the potential corresponds to the point  $(\phi_1=\arctan\sqrt{m_1/m_0},\;\phi_2=\arctan  \sqrt{m_2/(m_0+m_1))}$. 
The value of the
potential at this point is
$(m_0+m_1+m_2)^2$. Hence, the energy
$E$ is  bounded from below, i.e.
$E\ge (m_0+m_1+m_2)^2$.

As we mentioned before these results reflect
essentially the (1D)  $su(2)$-case. In fact,  all
systems we can construct using Cartan
subalgebras can be described in a unified
way as it was shown in \cite{Calzada1,winternitz,olmo99}.

\sect{A quantum  superintegrable $u(3)$-Hamiltonian system} 
\label{aquantumsuperintegrableu(3)hamiltoniansystem}

From the quantum point of view the Hamiltonian (\ref{hamiltonianou3}) takes the form
\begin{equation}\label{hh}
    H =  - \left(J_0^2 + J_1^2 + J_2^2 \right)
    +\frac{l_0^2-1/4}{(s_0)^2} + \frac{l_1^2-1/4}{(s_1)^2}+
    \frac{l_2^2-1/4}{(s_2)^2} ,
\end{equation}
 where $(l_0,l_1,l_2)\in \R^3$, and $J_i
=-\epsilon_{ijk}s_j\p_k$.

Also in the quantum case, the eigenvalue problem
$H\,\Phi = E\, \Phi$ 
after substituting the coordinates (\ref{coo}), takes the form of a
separable differential equation
\begin{equation}\label{h}
    \left[-\d(\phi_2)^2 +\tan(\phi_2) \d(\phi_2) +
    \frac{l_2^2{-}1/4}{\sin^{2}(\phi_2)} + \frac{1}{\cos^{2}(\phi_2)}
    \left( - \d(\phi_{1})^2 +    \frac{l_0^2{-}1/4}{\cos^{2}(\phi_1)} +
    \frac{l_1^2{-}1/4}{\sin^{2}(\phi_1)}\right)
    \right]\Phi = E\, \Phi .
\end{equation}
Taking the solutions separated in the variables $\phi_1$ and $\phi_2$ as 
${\bf  \Phi}(\phi_1,\phi_2) =f(\phi_1)g(\phi_2) $
 after replacing
in (\ref{h}) we get the equations 
\bea 
&&\left[ -
\d(\phi_{1})^2 +
    \frac{l_0^2-1/4}{\cos^{2}(\phi_1)} +
    \frac{l_1^2-1/4}{\sin^{2}(\phi_1)}\right] f(\phi_1) =
\alpha\,f(\phi_1),    \label{sisn21}\\[0.3cm]
&& \left[ -\d(\phi_2)^2 +
    \tan(\phi_2) \d(\phi_2) + \frac{\alpha}{\cos^{2}(\phi_2)}+
    \frac{l_2^2-1/4}{\sin^{2}(\phi_2)} \right]g(\phi_2) =
    E\,g(\phi_2) \,,\label{sisn22}
\eea 
where $\a$ is a separating constant. 

These two (one-variable dependent) equations  can be solved using the  standard factorizations obtaining 
polynomial solutions. Notice  that the results obtained for the
first equation will match in a certain way with those of the second
one giving rise to degenerate levels. 

\subsect{The factorization of the $\phi_1$-equation}

The 1D Hamiltonian corresponding to the equation  (\ref{sisn21}) in the variable
$\phi_1$ can be factorized using  the theory of factorizations by Infeld and Hull
\cite{infeld}.

The second order differential operator at the l.h.s.\ of equation\ (\ref{sisn21}) can
be written  as a product of first order operators 
\[
 H_{(0)}^{\phi_1}
= A^{+}_{0}\,A^{-}_{0} + \lambda_0 ,
 \] 
 where $A_{0}^{\pm} =\,\pm
\d(\phi_1) -  (l_0+1/2)\,\tan \phi_1 + (l_1+1/2)\,\cot \phi_1 $, and
$\lambda_0 \,= (l_0 + l_1+1)^2$. 
Also it is possible  to construct a family of operators
$\{A_{m}^{+},A_{m}^{-}\,,\lambda_{m}\,,H_{(m)}^{\phi_1}\}$, $m\in \Z$, 
where 
\bea 
&&A_{m}^{\pm} = \pm \d(\phi_1) -  (l_0 + m+1/2)\,\tan
\phi_1 + (l_1+m+1/2)\,\cot \phi_1  ,\label{aa}\\[0.3cm]
&&\lambda_{m} =  (l_0 + l_1 + 2\,m+1)^2 ,\nonumber \\[0.3cm]
&& H_{(m)}^{\phi_1}=  -\d(\phi_1)^2
    +\frac{ (l_{0}+m)^2 -1/4}{{\cos}^2 \phi_1 } +
    \frac{\,(l_{1}+m)^2-1/4}{{\sin}^2 \phi_1 }\,.   \label{hier}
\eea 
Hence, we obtain a  1D Hamiltonian hierarchy
(\ref{hier}), whose first element is  $H_{(0)}^{\phi_1}$,  satisfies
\be\label{co}
 H_{(m)}^{\phi_1} =
A^{+}_{m}\,A^{-}_{m} + \lambda_{m} = A^{-}_{m-1}\,A^{+}_{m-1} +
    \lambda_{m-1}.
\ee 
From it we see that the operators  $A_m^\pm$ are shape invariant intertwining operators, i.e.
\be\label{ah} 
A^{-}_{m}H_{(m)}^{\phi_1}=H_{(m+1)}^{\phi_1}A^{-}_{m}\
,\qquad A^{+}_{m}H_{(m+1)}^{\phi_1}=H_{(m)}^{\phi_1}A^{+}_{m}  .
\ee
Formally, the operators $A^\pm_{m}$ acting
on a Hamiltonian eigenfunction give another eigenfunction of a
consecutive Hamiltonian in the hierarchy with the same eigenvalue, i.e.
$$
A^{-}_{m} : {\cal H}_{m}^{\phi_1} \to {\cal H}_{m+1}^{\phi_1},\qquad
A^{+}_{m} : {\cal H}_{m+1}^{\phi_1} \to {\cal H}_{m}^{\phi_1}.
$$
where  ${\cal H}_{m}^{\phi_1}$ is  the eigenfunction space of $H_{(m)}^{\phi_1}$ .

The discrete spectrum and the physical eigenstates of
$H_{(0)}^{\phi_1}$  may be obtained, in principle,  from the fundamental states
$f_{(m)}^0$ (and their eigenvalues) of all the Hamiltonians of the
hierarchy $\{H_{(m)}^{\phi_1}\}$.  The  fundamental states are
determined by the equation
$
A^{-}_{m}\,f^{0}_{(m)} = 0
$ 
 whose solutions, up
to a normalization constant, are
 \be\label{eigenvector}
    f^{0}_{(m)}(\phi_1)=  \cos^{l_0 + m+1/2}(\phi_1)\,
    \sin^{l_1+m+1/2}(\phi_1)\,,
\ee 
with eigenvalues $\lambda_{m} = (l_0 + l_1 + 2\,m+1)^2$. 

The excited eigenfunction $f_{(0)}^{m}$ of $H_{(m)}^{\phi_1}$ can be obtained  from the ground eigenstate $f_{(m)}^0$ of $H_{(0)}^{\phi_1}$ (both with the same eigenvalue)  applying
consecutive operators  $A^+$ 
\be\begin{array}{lll}\label{excited}
    f_{(0)}^{m}& =& A^{+}_{0}\,A^{+}_{1}\cdots
A^{+}_{m-1}\,f_{(m)}^0 \\[0.3cm]
     &=&\,N\,   \sin^{l_1+1/2}(\phi_1)\,\,\cos^{l_0+1/2}(\phi_1)\,\,
       P_{m}^{(l_1,l_0)}[\cos(2\,\phi_1)] ,
\end{array}\ee
 where $P_{n}^{(a,b)}(x)$  are
Jacobi polynomials and $N$ a normalization constant. The
spectrum of the Hamiltonian  $H_{(0)}^{\phi_1}$ (\ref{sisn21}) is given
by 
\be\label{aespectro} 
\a = \lambda_{m}=(l_0 + l_1 +
2\,m+1)^2,\qquad m\in\Z^{\geq 0} \ .
\ee

\subsect{The dynamical algebras associated to the $\phi_1$-factorization}
\label{Thedynamicalalgebrasassociatedtothephi_1factorization}

The  shape invariant intertwining operators for the 1D Hamiltonian hierarchy $\{H_{(m)}^{\phi_1}\}$ determine some  Lie
algebras that we  can characterize as follows. 

Let us define free-index
operators $A^\pm$ acting inside the total space $\oplus_{m}{\cal H}_{m}$ 
from the above  $A^{\pm }_{(m)}$ by \cite{fernandez,refined00} 
\be
\label{aas}
\begin{array}{l}
A^+ f_{(m+1)}:= \frac12 A^+_{m} f_{(m+1)}\propto \tilde f_{(m)}\\[0.3cm]
A^- f_{(m)}:= \frac12 A^-_{m} f_{(m)} \propto \tilde f_{(m+1)}\\[0.3cm]
A \,f_{(m)}:= -\frac12(l_0 {+} l_1 {+} 2 m )f_{(m)}\propto f_{(m)}
\end{array}
\ee where $f_{(m)}$ (or $\tilde f_{(m)}$) denotes an eigenfunction
of $H_{(m)}$.  We can rewrite
(\ref{co}) as 
\be \label{com}
 [A,A^{\pm}] = \pm A^{\pm},\qquad [A^-,A^+] = - 2 A 
\ee 
assuming that the action is on any $f_{(m)}$. These commutators  determine
a Lie algebra $su(2)$ whose Casimir element is  
${\cal C}= A^+A^- +A(A-1)$. 

It can be proved (for more details see Ref.~\cite{olmo06}) that the  eigenstates of the Hamiltonians $H^{\phi_1}_{(m)}$ (\ref{eigenvector}) can be characterized,   if $l_0$ and $l_1$ are positive o zero integer numbers, in terms of the vectors of the  irreducible unitary representations (IUR) of $su(2)$, 
labeled by the parameter  $j$ such that  $2 j \in \Z^{\geq 0}$. Effectively, the ground states
$f_{(m)}^0$ of  $H^{\phi_1}_{(m)}$  are characterized by
\be\label{fu}
A^-f_{(m)}^0=0,\qquad
A\,f_{(m)}^0=-[(l_0+l_1+2m)/2] f_{(m)}^0 .
\ee
Then, we  identify (up to a normalization
constant)
\[
f_{(m)}^0 = |j_m,-j_m\rangle, \qquad j_m=(l_0+l_1+2m)/2 ,
\]
where $|j_,s\rangle$ denotes the  vectors of the IUR $`j'$ of $su(2)$.

The excited states of $H^{\phi_1}_{(m)}$ are obtained  using  
expression (\ref{excited}). So, the eigenstate of the $k$-th excited level of
$H^{\phi_1}_{(0)}$ is
\[
f_{(0)}^k\equiv |j_k+k,-j_k+k\rangle,\qquad j_k=(l_0+l_1+2k)/2, \qquad k=0,1,2\dots
\]
Moreover,  $H^{\phi_1}_{(0)}$ (as well as any $H^{\phi_1}_{(m)}$) can be expressed in 
terms of the $su(2)$-Casimir
$\cal C$ acting on such representations by
$H^{\phi_1}_{(0)}=4({\cal C} +1/4)$.
Hence, the eigenvalue equation for any of the excited states can be written
as follows
\[\begin {array}{lllr}
H^{\phi_1}_{(0)} f_{(0)}^k &\equiv & 4({\cal C}+1/4)|j_0+k,-j_0+k\rangle
= 4(j_0+k+1/2)^2
|j_0+k,-j_0+k\rangle &\\[1.ex] &= &(l_0+l_1 +2k+1)^2 f_{(0)}^k,& k=0,1,2,\dots 
\end{array}\]

However,  there is an ambiguity due  that different fundamental states (\ref{eigenvector}) with values of $l_0$ and $l_1$ 
 giving the same $j_0=(l_0+l_1)/2$ would lead to the same
$j_0$-representation of $su(2)$.

Adding the  diagonal operator $D$,
$D f_{(m)}:= (l_0-l_1) f_{(m)}$,
to the generators of  $su(2)$  (\ref{aas}) we obtain  $u(2)$. The eigenstates of the Hamiltonian hierarchy are now completly characterized by the IUR's of $u(2)$. 
 However, different 
$u(2)$-IUR's may give rise to (different) states with the same energy.  

On the other hand,  the states of the Hamiltonian hierarchies, when  
$l_0$ or $l_1$ are not in $\Z^{\geq 0}$,  correspond to  non-unitary representations of
$u(2)$, although they and their spectra are also given by formulae (\ref{eigenvector})  and (\ref{aespectro}), respectively.

From a classification  point of view it will be interesting to construct a new Lie algebra, obviously containing the subalgebra chain  $su(3) \subset u(3)$,  such that only one of
its IUR's characterizes all the eigenstates in the
hierarchy with same energy.

In order to build a such  dynamical algebra we need to
introduce  a a two-subindex notation, in terms of the parameters  $(l_0,l_1)$, for the intertwining operators.  
The  Hamiltonians  $H_{(m)}^{\phi_1}$ (\ref{hier}) will be
denoted by $H_{(l_0,l_1)}^{\phi_1}$, 
 its eigenfunctions  by $f_{(l_0,l_1)}$, 
and the factor operators $A_0^\pm$ in (\ref{aa}) will be
rewritten as $A_{(l_0,l_1)}^\pm$.
 In this way,  relation (\ref{ah}) can be expressed as 
\[\label{ahh}
A^{-}_{(l_0,l_1)}H_{(l_0,l_1)}^{\phi_1}=
H_{{(l_0+1,l_1+1)}}^{\phi_1}A^{-}_{(l_0,l_1)}\ ,\qquad
A^{+}_{(l_0,l_1)}H_{{(l_0+1,l_1+1)}}^{\phi_1}=
H_{{(l_0,l_1)}}^{\phi_1}A^{+}_{(l_0,l_1)}  .
\]
We can  also define the free-subindex operators $A^\pm,A, D$ as in
(\ref{aas}).

The fact that each two-parameter Hamiltonian
$H_{(l_0,l_1)}^{\phi_1}$ is invariant under the reflections
\[
I_0:(l_0,l_1)\to(-l_0,l_1),\qquad I_1:(l_0,l_1)\to(l_0,-l_1)
\]
originates  a second factorization (\cite{barut,quesne,dutt}) via conjugation of 
the operators $A^\pm,A, D$, 
\[
\begin{array}{lll}
I_0 A^\pm I_0=\tilde A^\pm, \qquad  &I_0 A I_0=\tilde A,
\qquad  &I_0 D I_0=\tilde D  ,\\[0.3cm]
\tilde I_1 A^\pm I_1=\tilde A^\mp, \qquad  &I_1 A I_1=-\tilde A,
\qquad  &I_1 D I_1=-\tilde D .
\end{array}
\] 
The explicit form of the new operators is
 \be\label{ta01}
  \tilde A_{(l_0,l_1)}^{\pm} = \pm
\d(\phi_1) +  (l_0{-}1/2) \,\tan \phi_1 +
       (l_1+1/2)\,\cot \phi_1,\qquad
\tilde A_{(l_0,l_1)}= -\frac12(-l_0 {+} l_1 ) .
 \ee 
 These operators $\{ \tilde A, \tilde A^{\pm}\}$ close a second Lie algebra
${su}(2)$, denoted $\widetilde{su}(2)$, which commutes with  the previous ${su}(2)$. Since, moreover,
 $D$ and $\tilde D$ essentially coincide with $\tilde A$ and $A$,
respectively, the new dynamical algebra is $su(2)\oplus \widetilde{su}(2)\approx so(4)$.

The action of the $so(4)$-generators on a Hamiltonian
$H_{(l_0,l_1)}^{\phi_1}$ originates a 2D parameter $so(4)$-hierarchy $\{ H_{l_0-n+m,l_1+n+m}\}$, $m,n\in\Z$, fixed by the initial values $(l_0,l_1)$.
Each energy level of this Hamiltonian hierarchy is degenerated and
the eigenstates are characterized by $so(4)$-representations.


\subsect{The factorization of the $\phi_2$-equation}

The  second equation (\ref{sisn22})  can be also factorized provided 
that the separation constant $\a$ is substituted by   the eigenvalues  obtained from the $\phi_1$-factorization $\a = \l_{m} = (l_0 + l_1 +2m)^2$.
The Hamiltonian associated to this $\phi_2$-equation (\ref{sisn22}) is
\be \label{2}
H_{(0)}^{\phi_2} = {\displaystyle -\p_{\phi_2}^2 +
\tan(\phi_2)\p_{\phi_2} + \frac{ (l_0 + l_1 +2m+1)^2}{
\cos^2(\phi_2) } +\frac{l_2^2-1/4}{\sin^2(\phi_2)} }.
\ee 
It can be factorized in terms of two first-order differential operators as 
$H_{(0)}^{\phi_2} = M^+_0 M^-_0 + \mu_0$. 
This Hamiltonian  $H_{(0)}^{\phi_2} $ is the first element of the Hamiltonian hierarchy
$\{H_{(n)}^{\phi_2}\},\; n\in \Z^{\geq 0}$, in the ${\phi_2}$--variable,  whose elements can be written as
\bea
&&  H_{(n)}^{\phi_2}
= M^+_n M^-_n + \mu_n = M^-_{n-1} M^+_{n-1} + \mu_{n-1} ,\nonumber\\[0.3cm]
&&M_n^\pm= \pm \p_{\phi_2} -
 (l_0 + l_1 +2(m+1) +n) \tan(\phi_2) + (l_2+n+1/2)\cot(\phi_2) ,\nonumber\\[0.3cm]
&& \mu_n =  (l_1 +l_0 +l_2 + 2n + 2m +3/2)(l_2+l_1 +l_0 +2n +2m+5/2) .\label{energia}
\eea 
The   energy  values $E_n$ are given by the factorization constant
 $\mu_n$, i.e.  $E = \mu_n$.
 The ground  states $g_{(n)}^{0}$ for this hierarchy are 
 \[
g_{(n)}^{0}(\phi_2) = N \cos(\phi_2)^{l_1 +l_0 +2m+1 } \sin(\phi_2)
^{l_2 +n+1/2 }\,.
\]
The eigenfunctions $g_{(0)}^{n}$ of the
initial Hamiltonian $H_{(0)}^{\phi_2}$ (\ref{2}) can be written as
\be\label{ef2}
g_{(0)}^{n}(\phi_2) = \cos(\phi_2)^{l_1+l_0+2m+1
}\sin(\phi_2)^{l_2+1/2}P_n^{(l_2+1/2,l_1 +l_0 +2m+1)}[\cos 2\phi_2].
\ee 
The  index-free operators
$M^\pm$, defined in a similar way as $A^\pm$ in (\ref{aas}),   close
again a Lie algebra $su(2)$.  The eigenfunctions
(\ref{ef2}) are square-integrable, but the $su(2)$-representations are
IUR provided that  the parameters $l_0,l_1,l_2$ belong to $\Z^{\geq 0}$.

The new factorization leads to a degeneration of
the energy levels indicating  that the underlying dynamical
symmetry could be larger than $so(4)$.

\begin{figure}[htp]
\centerline{
\includegraphics[scale=0.75]{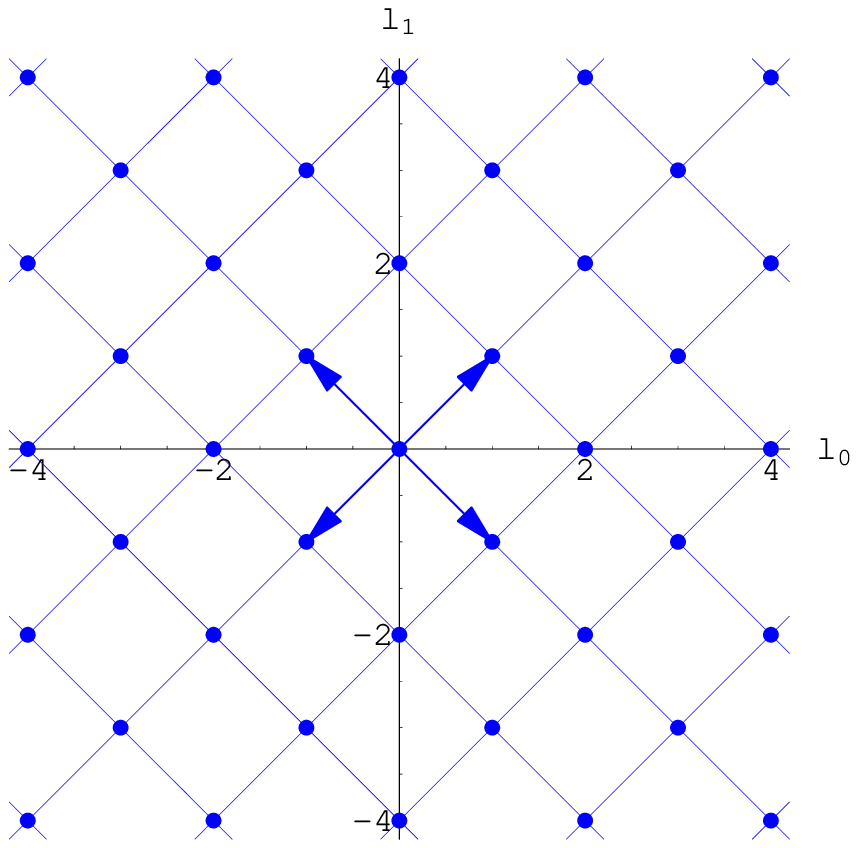}\hskip1cm
\includegraphics[scale=0.75]{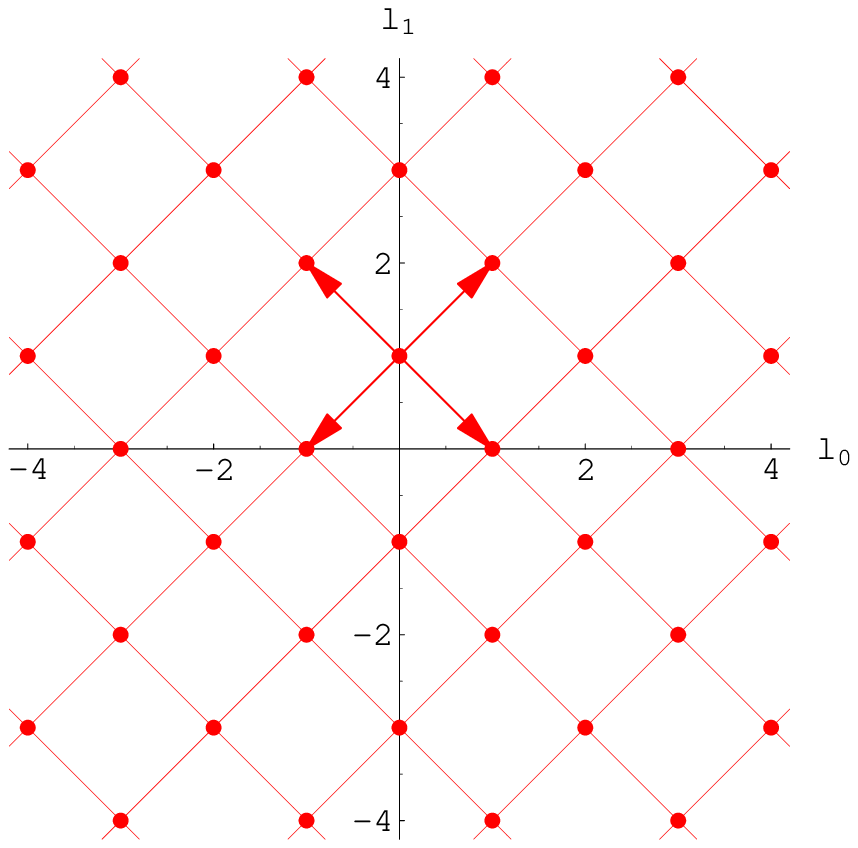}}
\caption{Plot of the two IUR's $so(4)$-hierarchies, where each point
represents a Hamiltonian. At the left it is the integer
$(l_0=0,l_1=0)$ and at the right the half-odd $(l_0=0,l_1=1)$
hierarchy. The arrows stand for the intertwining operators.}
\label{fig1}
\end{figure}

\begin{figure}[htp]
\centerline{
\includegraphics[scale=0.75]{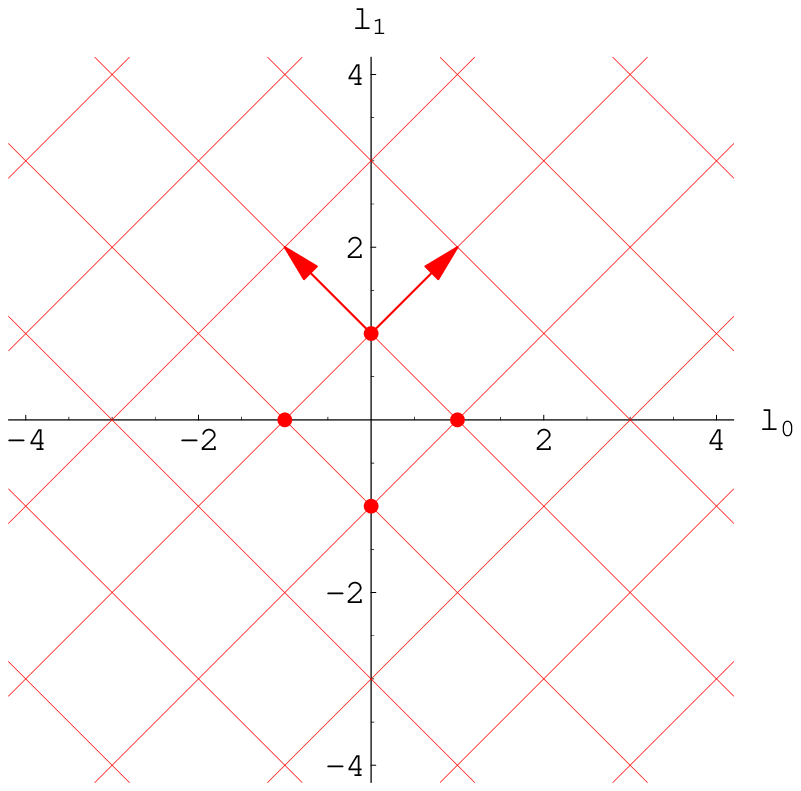}\hskip1cm
\includegraphics[scale=0.75]{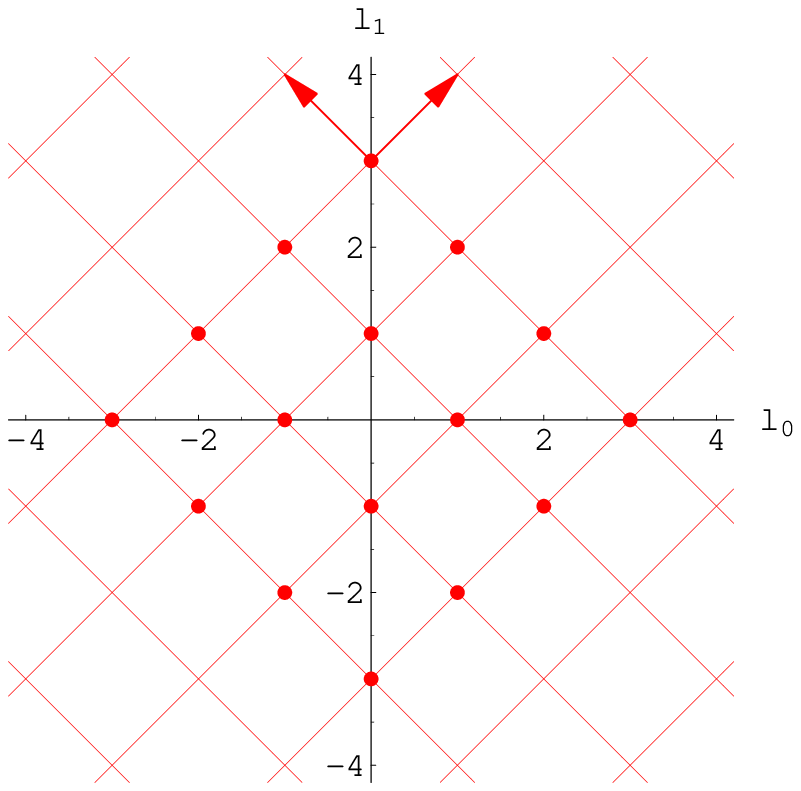}}
\caption{The points of the plots represent eigenstates of the
underlying half-odd Hamiltonian $so(4)$-hierarchy. At the left, the 4
eigenstates of the $\frac12\otimes \frac12$ $so(4)$-IUR that share the
energy $E=(1 +1)^2$, and at the right the 16 eigenstates of the
$\frac32\otimes \frac32$ $so(4)$-IUR with $E=(3 +1)^2$. } \label{fig2}
\end{figure}

Finally joining  both factorizations we obtain 
the square-integrable eigenfunctions separated in the variables
$(\phi_1,\phi_2)$ of the Hamiltonian (\ref{h})  
 \[\label{gs}
  \Phi_{m,n}(\phi_1,\phi_2) =
f^m_{(0)}(\phi_1)\, g^n_{(0)}(\phi_2),\qquad  m,n\in \Z^{\geq 0} ,
\]
where $f^m_{(0)}(\phi_1$ and $g^n_{(0)}(\phi_2)$ are given by the  expressions (\ref{excited}) and
(\ref{ef2}), respectively. Their corresponding eigenvalues $E_{n,m}$   (\ref{energia})
are degenerated for the  values of $m$ and $n$ whose sum $m+n$ keeps constant
\cite{pogosyan}  (see Figures~\ref{fig1} and~\ref{fig2}).

\sect{Dynamical symmetries of the   $u(3)$-Hamiltonian hierarchy}
\label{dynamicalsymmetries}

The spectrum of the $u(3)$-Hamiltonian system (\ref{hh}) suggestes a bigger
dynamical algebra of the Hamiltonian hierarchy.  By introducing three sets   
of intertwining operators closing an algebra $u(3)$  and using reflexion operators this algebra is 
enlarged to $so(6)$. These three sets of operators are related with  three set of spherical coordinates that we can take in the 2-sphere  submerged in a 3D ambient space with cartesian axes 
$\{s_0,s_1,s_2\}$.   Since the axes $(s_0,s_1,s_2)$ play a symmetric role in the
Hamiltonian  (\ref{hh}), we will take their cyclic rotations to get two other
intertwining sets. These two set of spherical coordinates also separate the Hamiltonian  (\ref{hh}).

\subsect{The $u(3)$-Hamiltonian hierarchies}

The Hamiltonian as well as the intertwining operators will be labelled now by  three parameters 
$(l_0,l_1,l_2)$ instead of two ones that we made used previously in 
section~\ref{Thedynamicalalgebrasassociatedtothephi_1factorization}.

The first set of sphere coordinates that we will use is that of coordinates $(\phi_1,\phi_2)$ such that the coordinates $\{s_0,s_1,s_2\}$ are given by (\ref{coo}).
The Hamiltonian (\ref{hh}) characterized by the parameters
$\ell\equiv (l_0,l_1,l_2)$ will be now denoted by
$H_{(l_0,l_1,l_2)}$, and the operators  (\ref{aa})
 will be rewritten  with a three-fold subindex
\[\label{als}
    A_{(l_0,l_1,l_2)}^{\pm}= \pm \d(\phi_1) -  (l_0+1/2)
    \,\tan \phi_1 + (l_1+1/2)\,\cot \phi_1 .
\]
The differential
operators (\ref{aa}) depend only on the variable
$\phi_1$, hence they do not act on  the second separating variable $\phi_2$. So,
we have the intertwining
relations 
\[ 
A^{-}_{(l_0,l_1,l_2)}H_{(l_0,l_1,l_2)}=
H_{(l_0+1,l_1+1,l_2)}A^{-}_{(l_0,l_1,l_2)}, \qquad
A^{+}_{(l_0,l_1,l_2)}H_{(l_0+1,l_1+1,l_2)}=H_{(l_0,l_1,l_2)}A^{+}_{(l_0,l_1,l_2)} .
\]
The global operators acting on eigenfunctions of these
$H_{(l_0,l_1,l_2)}$ defined    as in (\ref{aas}),
\bea 
&&A^+ \Phi_{(l_0+1,l_1+1,l_2)}:=
\frac12 A^+_{(l_0,l_1,l_2)} \Phi_{(l_0+1,l_1+1,l_2)} \propto \tilde
\Phi_{(l_0,l_1,l_2)},\nonumber\\
&&A^- \Phi_{(l_0,l_1,l_2)}:= \frac12 A^-_{(l_0,l_1,l_2)}
\Phi_{(l_0,l_1,l_2)} \propto \tilde \Phi_{(l_0+1,l_1+1,l_2)},
\nonumber\label{mas}\\
&&A\, \Phi_{(l_0,l_1,l_2)}:= -\frac12(l_0 +
l_1)\Phi_{(l_0,l_1,l_2)}, \nonumber
 \eea 
close an algebra  $su(2)$ with  commutators like  (\ref{com}). Notice 
that now these operators are acting on the total wavefunction
of the complete Hamiltonians like $H_{(l_0,l_1,l_2)}$, not just on a
factor function in one-variable. 

 We will take the spherical
coordinates choosing as `third axis'  $s_1$ instead of $s_2$,
\[\label{coo2}
   s_2 = \cos(\xi_2)\,\cos(\xi_1),\qquad
   s_0 = \cos(\xi_2)\, \sin(\xi_1),\qquad
   s_1 \,= \sin(\xi_2)\, .
\] 
The corresponding  intertwining operators
$B_{(l_0,l_1,l_2)}^{\pm}$ are defined in a similar  way to $A_{(l_0,l_1,l_2)}^{\pm}$. The explicit
 expressions for the new set in terms of the initial coordinates (\ref{coo}) are
\[\label{bls}
 B_{(l_0,l_1,l_2)}^{\pm}= \pm (\sin\phi_1\tan
\phi_2\d(\phi_1) +\cos\phi_1\d(\phi_2)) -  (l_2{+}1/2)
\,\cos\phi_1{\rm cot}\phi_2 + (l_0{+}1/2)\,{\rm sec}\phi_1{\rm
tan}\phi_2 .
 \] 
Their intertwining action on  the  Hamiltonians is 
\[ 
B^{-}_{(l_0,l_1,l_2)}H_{(l_0,l_1,l_2)}=
H_{(l_0+1,l_1,l_2+1)}B^{-}_{(l_0,l_1,l_2)},\qquad
B^{+}_{(l_0,l_1,l_2)}H_{(l_0+1,l_1,l_2+1)}=H_{(l_0,l_1,l_2)}B^{+}_{(l_0,l_1,l_2)} .
\]
 The `global' operators, defined by means of multiplicative
constant, spann also an  algebra $su(2)$.

The spherical
coordinates around the $s_0$ axis are
 \[\label{coo3}
   s_1 = \cos \theta_2 \,\cos \theta_1 ,\qquad
   s_2 = \cos \theta_2 \, \sin \theta_1 ,\qquad
   s_0 \,= \sin(\theta_2).
\]
We obtain a new pair of operators,
that written in terms of the original  variables $(\phi_1,\phi_2)$ are 
\[\label{cls}
C_{(l_0,l_1,l_2)}^{\pm}= \pm (\cos\phi_1\tan \phi_2\d(\phi_1)
-\sin\phi_1\d(\phi_2)) +  (l_1{-}1/2) \,{\rm cosec}\phi_1{\rm
tan}\phi_2 + (l_2{+}1/2)\,{\rm sin}\phi_1{\rm cot}\phi_2  .
\] 
These operators  intertwin  the Hamiltonians in the following way
 \[
 C^{-}_{(l_0,l_1,l_2)}H_{(l_0,l_1,l_2)}=
H_{(l_0,l_1-1,l_2+1)}C^{-}_{(l_0,l_1,l_2)},\qquad
C^{+}_{(l_0,l_1,l_2)}H_{(l_0,l_1-1,l_2+1)}=H_{(l_0,l_1,l_2)}C^{+}_{(l_0,l_1,l_2)} .
\] 
The `global' operators (again $1/2$ times the 'old' ones) close
the third $su(2)$.

All these transformations  $\{A^{\pm}, A, B^{\pm}, B, C^{\pm}, C\}$ ($C=B-A$) spann  an  algebra $su(3)$. 
The $su(3)$-Casimir operator  is given by
\[\label{cas} 
{\cal C}= A^+A^- + B^+B^- + C^+C^- + \frac23 A(A-3/2)
+ \frac23 B(B-3/2) + \frac23 C(C-3/2) .
 \] 
We obtain  $u(3)$ by adding the central diagonal operator $D:=l_0-l_1-l_2$.

The global
operator convention can be adopted for the Hamiltonians $H$  in the $u(3)$-hierarchy by
defining its action on the eigenfunctions $\Phi_{(l_1,l_2,l_3)}$ of
$H_{(l_1,l_2,l_3)}$ by 
\[ 
H\Phi_{(l_1,l_2,l_3)}:= H_{(l_1,l_2,l_3)}
\Phi_{(l_1,l_2,l_3)} .
\] 
Then, the Hamiltonian
$H$   can be expressed in terms of both Casimir operators, 
\be\label{esp} H= 4\,{\cal
C}-\frac13\, D^2+\frac{15}4 .
 \ee 
 Hence, the  Hamiltonian can be written as a certain quadratic function
of the operators $\{A^\pm,B^\pm,C^\pm\}$ generalizing the usual
factorization for 1D systems.

In this way  we have built an  algebra $u(3)$ of intertwining
operators that, once fixed the initial Hamiltonian with parameter
values $(l_0,l_1,l_2)$, generate  a two-parameter Hamiltonian
hierarchy
\[
 \{H_{(l_0+m,l_1+m-n,l_2+n)} \},\qquad
m,n\in\Z ,
\]
where the points $(l_0+m,l_1+m-n,l_2+n)$ lie on a certain plane
$D=d_0$.

One can prove that the eigenstates of this Hamiltonian hierachy 
are connected to the IUR's of $u(3)$. Fundamental states $\Phi$ 
annihilated by $A^-$ and
$C^-$ (simple roots of $su(3)$) 
\[\label{ac} 
A_\ell^-\Phi_\ell
= C_\ell^-\Phi_\ell =0 
\]
only  exist  when $l_1=0$, 
\[\label{ground} 
\Phi_\ell(\phi_1,\phi_2)= N
\,\cos^{l_0{+}1/2}(\phi_1)\sin^{1/2}(\phi_1)\cos^{l_0+1}(\phi_2)\sin^{l_2{+}1/2}(\phi_2) ,
\] 
whit $N$  a normalizing constant. The diagonal operators act
on them  as 
\be\label{mn}
\begin{array}{lll}
A\, \Phi_\ell = -l_0/2\, \Phi_\ell,\quad & l_0 = m,\qquad l_1=0,\qquad & m=0,1,2,\dots\\[0.3cm]
C\, \Phi_\ell = -l_2/2\, \Phi_\ell,\quad & l_2 = n,\qquad & n=0,1,2,\dots
\end{array}
\ee
This shows that $\Phi_\ell$ is the lowest state of the
IUR $j_1=m/2$ of the  subalgebra $su(2)$ generated by $\{A^\pm,A\}$,
and of the IUR $j_2=n/2$ of the subalgebra $su(2)$ closed by
$\{C^\pm,C\}$. Such a $su(3)$-representation  will be denoted
$(m,n)$, $m,n\in\Z^{\geq 0}$. The points (labelling the states) of this
representation obtained from $\Phi_\ell$   lie on the plane
$D=m-n$ inside the $\ell$-parameter space.

The energy for the states of the IUR determined by the lowest
state (\ref{mn}) with
  parameters $(l_0,0,l_2)$,  
is given (\ref{esp}) by 
\be\label{ee} 
E=(l_0 + l_2 + 3/2) ( l_0  + l_2 + 5/2)=(m
+ n + 3/2) ( m  + n + 5/2) . \ee 
Note that the IUR's labelled by $(m,n)$
with the same value $m+n$ are associated to states with the same energy.
We call such IUR's a iso-energy series. This degeneration will be broken  using the 
 algebra  $so(6)$.

\subsection{The $so(6)$-hierarchy}

Making use of some relevant discrete symmetries, following the procedure of 
section~\ref{Thedynamicalalgebrasassociatedtothephi_1factorization}, 
the dynamical algebra $u(3)$ can be enlarged to $so(6)$.

The
Hamiltonian $H_{(l_0,l_1,l_2)}$  (\ref{hh}) is invariant under reflections in
the parameter space $(l_0,l_1,l_2)$
 \[
I_0:(l_0,l_1,l_2)\to (-l_0,l_1,l_2),\qquad 
I_1:(l_0,l_1,l_2)\to (l_0,-l_1,l_2),\qquad 
I_2:(l_0,l_1,l_2)\to (l_0,l_1,-l_2)
\]
These symmetries, $I_i$, can be directly implemented in
the eigenfunction space, giving by  conjugation  
another set
of intertwining operators (${}_i\!X = I_i \, X\,  I_i,\;   i=0,1,2$)
  closing an isomorphic Lie algebra
${}_iu(3)$.
They are (now labelled with a tilde)
\[
\begin{array}{ll}
\{A^\pm,B^\pm,C^\pm\} \stackrel{I_0} \longrightarrow  &\{\tilde
A^\mp,\tilde B^\mp,C^\pm\},\\[2.ex]
\{A^\pm,B^\pm,C^\pm\} \stackrel{I_1} \longrightarrow  &\{\tilde
A^\pm, B^\pm,\tilde C^\pm\},\\[2.ex]
\{A^\pm,B^\pm,C^\pm\} \stackrel{I_2} \longrightarrow  &
\{A^\pm,\tilde B^\pm, \tilde C^\mp\} .
\end{array}\]
For instance, the sets $\{A^\pm,A\}$ and $\{\tilde
A^\pm,\tilde A\}$ close the two commuting $su(2)$  of
section~\ref{Thedynamicalalgebrasassociatedtothephi_1factorization}. 

The explicit expression for these new intertwining operators  
can be easily obtained in the same way as
was done in (\ref{ta01}). They close a Lie algebra of rank 3, $so(6)$. Instead of the six non-independent  generators $A,\tilde A, B,\tilde B, C,\tilde C$   it is enough to consider three independent
diagonal operators $L_0,L_1,L_2$ defined by 
$ L_i\,
\Psi_{(l_0,l_1,l_2)} := l_i \,\Psi_{(l_0,l_1,l_2)}$.
 The Hamiltonian can be expressed in terms of the $so(6)$-Casimir operator  
  by means of the `symmetrization' of the $u(3)$-Hamiltonian (\ref{esp})
  \[\label{CAS}
H =\{A^+,A^-\}+\{B^+,B^-\}+\{C^+,C^-\}
+ \{\tilde A^+, \tilde A^-\}+\{\tilde B^+, \tilde B^-\}+ \{\tilde
C^+, \tilde C^-\}+ {L_0}^2+{L_1}^2+{L_2}^2 +\frac{41}{12}.
\]

The intertwining generators of $so(6)$ give rise to larger
3D Hamiltonian hierarchies
\[
\{H_{(l_0+m+p,l_1+m-n-p,l_2+n)} \},\qquad m,n,p\in\Z ,
\]
each one including a class of the previous ones coming from $u(3)$.
The eigenstates of these $so(6)$-hierarchies can be classified
in terms of $so(6)$ representations.


\section*{Acknowledgments}
This work has been partially supported by DGES of the
Ministerio de Educaci\'on y Ciencia of Spain under Projects BMF2002-02000 and FIS2005-03989
and Junta de Castilla y Le\'on (Spain) (Project VA013C05).



\begin{thebibliography}{20}
   
 \bibitem{Morse} P.M. Morse, {\it Phys. Rev}. {\bf 34}  (1929) 57.
   
 \bibitem{Posch} G. Posch and E. Teller, {\it Z. Phys}. {\bf 83} (1933) 143.
  
 \bibitem{smorodinski}
P. Winternitz, A. Smorodinsky, M. Uhlir and J. Fris, {\it Soviet J. Nuclear Phys}. {\bf 4} (1967) 444 .

\bibitem{Evans1} N.W. Evans, Phys. {\it Phys. Rev}. {\bf 41A} (1990)  5666;
     {\it Phys. Lett}.  {\bf 147A}  (1990) 483; {\it J. Math. Phys}. {\bf 32} (1991)  3369.

\bibitem{Calogero1} F. Calogero, {\it J. Math. Phys}. {\bf 10} (1969)  2191;
     {\it J. Math. Phys}. {\bf 10} (1969)  2197.
 
\bibitem{Sutherland} B. Sutherland, {\it Phys. Rev.}  {\bf 4A} (1971)  2019.
   
\bibitem{infeld} L. Infeld and T.E. Hull,  {\it Rev. Mod. Phys}.  {\bf 23} (1951) 21.

\bibitem{Eisenhart1} L.P. Eisenhart, {\it Ann. Math.} {\bf 35}   (1934)  284.
  
\bibitem{Marsden} J.~Marsden and A.~Weinstein,
{\it Rep.~Math.~Phys}. {\bf 5} (1974) 121.

\bibitem{Grabowski} J. Grabowski, G. Landi, G. Marmo and G. Vilasi,
        {\it Fortschr. Phys}. {\bf 42} (1996)  393.
   
   \bibitem{kuru1}
  \c{S}. Kuru, A. Te\v{g}men and A. Ver\c{c}in,
{\it  J. Math. Phys}. {\bf 42} (2001)  3344.

\bibitem{kuru2}
B. Demircio\v{g}lu,  \c{S}. Kuru, M. \"{O}nder and A. Ver\c{c}in,
 {\it  J. Math. Phys}. {\bf 43} (2002)  2133.

 \bibitem{samani}
K.A. Samani and  M. Zarei,
{\it Ann. Phys}. {\bf 316} (2005)  466.

\bibitem{ranada}
 M.F.~Ra\~nada, {\it J. Math. Phys}. {\bf 36} (1995) 3541;
 {\bf 38} (1997) 4165;
 {\bf 40} (1999) 236;
 {\bf 41} (2000) 2121.
 
\bibitem{santander}
 M.F.~Ra\~nada and M.~Santander, {\it  J. Math. Phys}. {\bf 40} (1999)
    5026;   {\bf 43} (2002)  431;  {\bf 44} (2003)  2149.

\bibitem{ioffe}
F. Cannata, M.V. Ioffe and D.N. Nishnianidze, {\it J. Phys.  A} {\bf 35}  (2002) 1389.

\bibitem{Kobayashi} S. Kobayashi and K. Nomizu, {\it Foundations
        of Differential Geometry} (Wiley, New York, 1969).

\bibitem{Calzada1} J.A. Calzada, M.A. del Olmo, M.A. Rodr\'{\i}guez,
        {\it J. Geom. Phys}. {\bf 23} (1997)  14.
   
\bibitem{Olmo1} M.A. del Olmo, M.A. Rodr\'{\i}guez, P. Winternitz and
        H. Zassenhaus, {\it Linear Algebr. Appl}. {\bf 135} (1990) 79.
   
\bibitem{winternitz} M.A. del Olmo, M.A. Rodr\'{\i}guez and P. Winternitz,
{\it J. Math. Phys}. {\bf 34} (1993) 5118; {\it Fortschritte der Physik} {\bf 44}  (1996) 91. 

\bibitem{olmo99} J.A. Calzada, M.A. del Olmo and M.A. Rodr\'{\i}guez,
{\it J. Math. Phys}. {\bf 40} (1999) 88.

\bibitem{Boyer}C.P. Boyer, E.G. Kalnins and P. Winternitz, SIAM
     {\it J. Appl. Math}.  {\bf 16} (1985) 93.
  
\bibitem{fernandez}D.J. Fern\'andez , J. Negro and M.A. del~Olmo,
        {\it Ann.  Phys}. {\bf 252} (1996)  386.
    
\bibitem{refined00} J. Negro, L.M. Nieto and O. Rosas-Ortiz,
       {\it  J. Phys. A}  {\bf 33} (2000) 7207.

\bibitem {olmo06}  J.A. Calzada, J. Negro and M.A. del Olmo, ``Superintegrable quantum $u(3)$-systems and higher rank factorizations''. math-ph/0601067.

 \bibitem{pogosyan} E.G. Kalnins, W. Miller, G.S. Pogosyan,
        {\it J. Math. Phys}. {\bf 37}  (1996) 6439.

\bibitem{barut}
A.O. Barut, A. Inomata and R. Wilson, {\it J. Phys. A} {\bf 20} 
(1987) 4075;  {\it J. Phys. A} {\bf 20}  (1987) 4083.

\bibitem{quesne}
A. del Sol Mesa, C. Quesne and Yu F. Smirnov,  {\it J. Phys. A} {\bf 31} 
(1998) 321.

\bibitem{dutt}
M. Dutt, A. Gangopadhyaya, C. Rosinaru and U. Sukhatme,  {\it J. Phys. A}   {\bf 34} 
(2001) 4129.

\end{thebibliography}
\end{document}